\documentclass[final,12pt]{article}

\usepackage{graphicx}
\usepackage{url}
\usepackage[a4paper,twoside]{geometry}

\begin{document}

\vspace*{4ex}
\begin{center}
{\Large \bf 
Fitting DVCS amplitude in moment-space\\[1ex]
approach to GPDs%
\footnotetext{Talks given by K.K.~at DIS 2008, 7--11 April 2008,
London \cite{Kumericki:DIS2008slides} and D.M.~at International
Workshop on Diffraction in High-Energy Physics, 9--14 September
2008, La Londe-les-Maures, France \cite{Mueller:DIF2008slides}.}
}

\vspace*{4ex}
{\bf
K.~Kumeri{\v c}ki$^{a}$, 
D.~M\"uller$^{b}$, and
K.~Passek-Kumeri\v{c}ki$^{c}$}

\vspace*{4ex} 
{\it $^a$Department of Physics, Faculty of Science, University of Zagreb\\
 P.O.B. 331, HR-10002 Zagreb, Croatia}

\vspace{1ex} 
{\it $^b$Institut f\"ur Theoretische Physik II, Ruhr-Universit\"at Bochum\\
 D-44780 Bochum, Germany}

\vspace{1ex} 
{\it $^c$Theoretical Physics Division, Rudjer Bo{\v s}kovi{\'c} Institute\\
 P.O.Box 180, HR-10002 Zagreb, Croatia}

\vspace{4ex}
\noindent
\begin{minipage}{28cc}
{\bf Abstract:}
We describe small-$x_{\rm Bj}$ deeply virtual Compton scattering
measurements at HERA in terms of generalized parton distributions
at leading order of perturbation series.
\end{minipage}

\end{center}

\vspace{3ex}

\vspace{1ex}

\vspace{4ex}

\section{Introduction}

Deeply virtual Compton scattering (DVCS), $\gamma^{*}(q_1)\,
p(P_1) \to \gamma(q_2)\, p(P_2)$, is viewed as the cleanest
process to access generalized parton distributions (GPDs), which
encode a partonic description of the nucleon,
cf.~Refs.~\cite{Diehl:2003ny,Belitsky:2005qn}. In the
kinematics of H1 and ZEUS collider experiments at HERA, the DVCS
cross section is to a large extent dominated by the flavor singlet
part of the helicity conserved Compton form factor (CFF) ${}^{\rm
S}\!\mathcal{H}$:
\begin{eqnarray}
\label{Def-CroSec1}
\frac{d\sigma}{d t}(W,t,{\cal Q}^2) \approx
\frac{4   \pi \alpha^2 }{{\cal Q}^4}\xi^2
\left| {}^{\rm S}\!\mathcal{H}\left(\xi,t=\Delta^2,{\cal Q}^2\right) \right|^2
\Big|_{\xi = {\cal Q}^2 / (2 W^2+{\cal Q}^2)}\,.
\end{eqnarray}
Here, $W$ is the c.o.m.~energy, $\Delta=P_2 - P_1$ is the momentum
transfer, $-\mathcal{Q}^2=q_{1}^2$ is the incident photon
virtuality, and $\xi\approx x_{\rm Bj}/2$ is a Bjorken-like
scaling variable.

The CFF ${}^{\rm S}\!\mathcal{H}$ factorizes further into a
convolution of the partonic, i.e., hard scattering, amplitude
$\mbox{\boldmath $C$}=({}^{\Sigma}C, {}^{G}\! C)$ and GPDs
$\mbox{\boldmath $H$}=({}^{\Sigma}H, {}^{G}\! H)$,
($\Sigma$=singlet quark, $G$=gluon),
\begin{equation}
{}^{\rm S}\!\mathcal{H}(\xi, t, \mathcal{Q}^2) =
\int_{-1}^1\! {\rm d}x\; \mbox{\boldmath $C$} (x, \xi, \mathcal{Q}^2/\mu^2,
\alpha_{s}(\mu)) \;
\mbox{\boldmath $H$} (x, \eta=\xi, t, \mu^2) \; ,
\label{eq:conv}
\end{equation}
where the skewness parameter $\eta=-\Delta\cdot q/(P_1+P_2)\cdot
q$ is set equal to $\xi$. The factorization scale $\mu$ separates
short- and long-distance dynamics and is often taken as
$\mu=\mathcal{Q}$. The scale dependence is governed by evolution
equations. They also tell us that the evolution effect depends on
the $(x,\eta)$ GPD shape itself. Note that gluons do not directly
enter the DVCS amplitude at leading order (LO), but rather drive
the evolution of singlet quarks.

Since  the momentum fraction $x$ is integrated out in the
amplitude (\ref{eq:conv}), GPDs cannot be directly revealed.
In the quest for a realistic GPD model, GPD values along the two
trajectories $\eta=0$ and $\eta=x$ play a prominent role%
\footnote{These trajectories appear in a GPD \emph{sum rule
family}, which can serve as powerful modelling tool
\cite{Kumericki:2008di}. }. Namely,  GPDs on these trajectories
are given at LO by  DIS structure function and imaginary part of
DVCS amplitude, respectively. In this sense they are
experimentally measurable. Thus, a modelling strategy  that places
emphasis on these trajectories has a good chance to be efficient
in a global DVCS fit and to capture the physical GPD content,
giving us a link to a partonic interpretation.

It is the objective of this work to find GPDs that satisfy the
well-known theoretical constraints
\cite{Diehl:2003ny,Belitsky:2005qn} and provide a good fit to all
available H1 and ZEUS DVCS data. Our particular concern is the LO
description, where so far this goal has not been reached. Since
the real part of the CFF $\cal H$ can be calculated from a
dispersion relation, our outcome might be relevant for the real
part in fixed target kinematics, too. Beyond LO these DVCS data
can be described within specific GPD models, see, e.g.,
Refs.~\cite{FreMcD01a,Kumericki:2007sa}.

\section{Conformal moment-space approach}

It is convenient to work with conformal GPD moments.  For integral
conformal spin $j+2$ they are defined by  convolution with
Gegenbauer polynomials $C_j^{\nu} (x)$, e.g., for quarks:
\begin{equation}
H^{q}_{j}(\eta, t, \mu^2) \equiv \frac{\scriptstyle
\Gamma(3/2)\Gamma(j+1)}{\scriptstyle 2^{j+1} \Gamma(j+3/2)}\int_{-1}^1\!
{\rm d}x\; \eta^j\,
 C_j^{3/2} (x/\eta)\, H^{q}(x, \eta,  t, \mu^2) \;,
\label{eq:defconf}
\end{equation}
where $q$ is the flavor index. The normalization ensures that in
the forward limit $\Delta\to 0$ the conformal moments simply
reduce to familiar Mellin moments of parton distribution functions
(PDFs). Some of the advantages of working with conformal moments
are
\begin{itemize}
\item Conformal moments evolve autonomously at LO and, in a special
factorization scheme, even at next-to-leading order (NLO).
\item Powerful analytic methods of complex $j$ plane are available (similar
to complex angular momentum). This also allows to build a stable
and fast routine for fitting.
\item New possibilities for {GPD modelling}, which makes direct contact to
the $t$-channel SO(3) partial wave expansion, Regge phenomenology, and lattice
measurements.
\end{itemize}
In moment space the convolution formula (\ref{eq:conv}) yields formally a
divergent series
over the integral conformal spin. Analogously to a SO(3) partial wave expansion,
it can be resummmed by means of a Mellin-Barnes integral representation
\cite{MueSch05,Kumericki:2006xx,Kumericki:2007sa}
 \begin{equation}
 \label{Def-MB-int}
 {^{\rm S}\!{\cal H}}(\xi,t,{\cal Q}^2)
 = \frac{1}{2i}\int_{c-i \infty}^{c+ i \infty}\!
 dj\,\xi^{-j-1} \left[i +
 \tan
 \left(\frac{\pi j}{2}\right) \right]
 \mbox{\boldmath
 $C$}_{j}({\cal Q}^2/\mu^2,\alpha_s(\mu))
 \mbox{\boldmath $H$}_{j}(\xi,t,\mu^2)
\, .
\end{equation}

GPD moments $H_j$, cf.~Eq.~(\ref{eq:defconf}), can  be decomposed
in $t$-channel SO(3) partial waves, i.e., Wigner matrices
$d^{J}_{0,\nu}(\cos\theta)$,  which are labelled by angular
momentum $J$ and hadron helicity differences ($\nu=0,\pm 1$)
\cite{Polyakov:2002wz,Diehl:2003ny,Kumericki:2007sa}. The cosine
of the scattering angle $\theta$ might be approximated by
$-1/\eta$. We include an effective leading Regge pole at
$\alpha(t)=\alpha_0 + \alpha' t$ in the partial wave amplitudes
and parameterize their residual $t$-dependence by a dipole with
cut-off mass $M_J$ (or, alternatively, with an exponential $t$-dependence).
Our model (for assignments see
Ref.~\cite{Diehl:2003ny}) is given at a input scale ${\cal Q}_0$ and
reads for integral $j$:
\begin{equation}
H_{j}(\eta, t, {\cal Q}_0^2) = \sum_{J=J^{\rm min}}^{j+1} h_{j}^J \frac{1}{J - \alpha(t)}
\frac{1}{\left(1-\frac{t}{M_J^{2}}\right)^2} \: \eta^{j+1-J} d^{J}_{0,\nu}(-1/\eta) \;,
\label{eq:so3}
\end{equation}
where $h^J_{j}$ are strengths of partial waves. The PDF Mellin
moments $h^{J=j+1}_{j}/(j+1-\alpha_0)$  are fixed from DIS, the
remaining parameters are constrained by DVCS fits.

In practice the parameter space in the ansatz (\ref{eq:so3}) must
be further reduced. We employ three GPD models: i.~taking only the
leading $J=j+1$ SO(3) partial wave (l-PW), ii.~including the
next-leading $J=j-1$ one (nl-PW), and iii.~performing a model
dependent resummation of SO(3) partial waves ($\Sigma$-PW). To set
up the third model, we assume that the GPDs in the vicinity of
$x=\eta$ behave as $\frac{1}{\eta}
(\frac{x+\eta}{1+\eta})^{1-\alpha(t)}$. Relevant for us
is the small $\eta$ behavior of the resulting conformal moments
for complex-valued $j$ \cite{MueSch05}, which is given by a series
of `conformal' daughter poles $j=\alpha(t)-1-n$ with
$n=2,4,6,\cdots.$ All of them contribute to the leading Regge
behavior of the CFF ${\cal H}$. The strength of the non-leading
SO(3) partial waves in the nl- and $\Sigma$-PW model for sea
quarks and gluons is controlled by skewness parameters. These
parameters allow us to adjust the normalization of CFFs. Note that
Mellin-Barnes and ``dual'' \cite{Polyakov:2002wz} GPD
parameterizations are related. In the latter $\rho=j+1-J$ is taken
as a parameter and a `forward-like' momentum fraction $z$-integral
replaces the $j$-integral (\ref{Def-MB-int}). Our l- and nl-PW
models are equivalent to a minimalist ($\rho=0$) and a minimal
($\rho=\{0,2\}$) ``dual'' model, respectively \cite{Pol07}.

\section{Results}

\begin{figure}
\centerline{\includegraphics[scale=0.51,angle=-90]{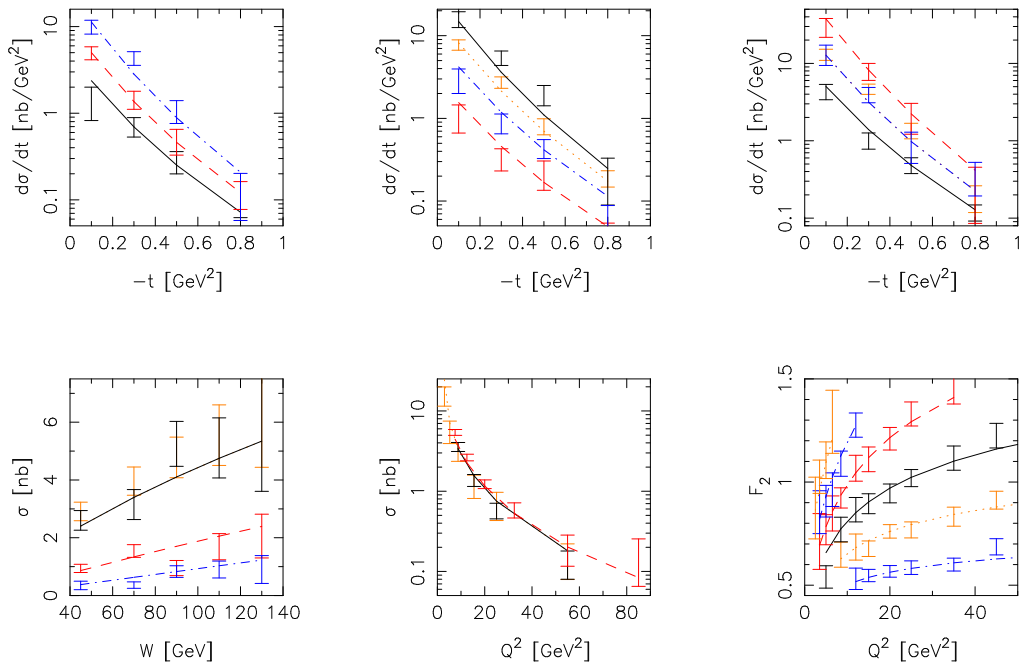}}
\caption{\label{fig:fit} Fit of $\Sigma$-PW model at LO to DVCS and DIS data from
\cite{:2007cz,Aktas:2005ty,Chekanov:2003ya,Aid:1996au}. }
\end{figure}
The l-PW model is the simplest one with a rigid normalization,
i.e., $\cal H$ at ${\cal Q}_0$ and $t=0$ is given by the
unpolarized sea quark PDF enhanced by the Clebsch-Gordan
coefficient
\begin{equation}
\label{CG-coe}
\frac{2^{1+\alpha_0}
\Gamma(\alpha_0+3/2)}{\sqrt{\pi}\Gamma(\alpha_0+2)} \sim 1.5.
\end{equation}
Within such a simplified model the $t$-slope is poorly described
at LO, since it is used to adjust the normalization, whereas at NLO
or beyond  good fits are obtained \cite{Kumericki:2007sa}.  To
have a  LO description, one has to include non-leading SO(3) partial
waves. Both of our flexible GPD models (nl-PW and $\Sigma$-PW) allow for good DVCS fits at
LO, illustrated for the $\Sigma$-PW model in
Fig.~\ref{fig:fit}, and beyond.  The $t$-dependence is well
described with $\alpha^\prime = 0.15/{\rm GeV}^2$ and $M_j \sim
0.6\, {\rm GeV}$, taken at the initial scale ${\cal Q}_0 =
2$ GeV. Note that to some extent these parameters are correlated.
The last panel shows fit to DIS structure function $F_2$, which
fixes the normalization $h^{j+1}_{j}$ of leading SO(3)
partial wave and  intercept $\alpha_0$. We add that gluonic GPD is
not well constrained, e.g., its $t$-slope can be taken from
$J/\psi$ data \cite{StrWei03}.

\begin{figure}[t]
\centerline{
\includegraphics[scale=0.6,clip]{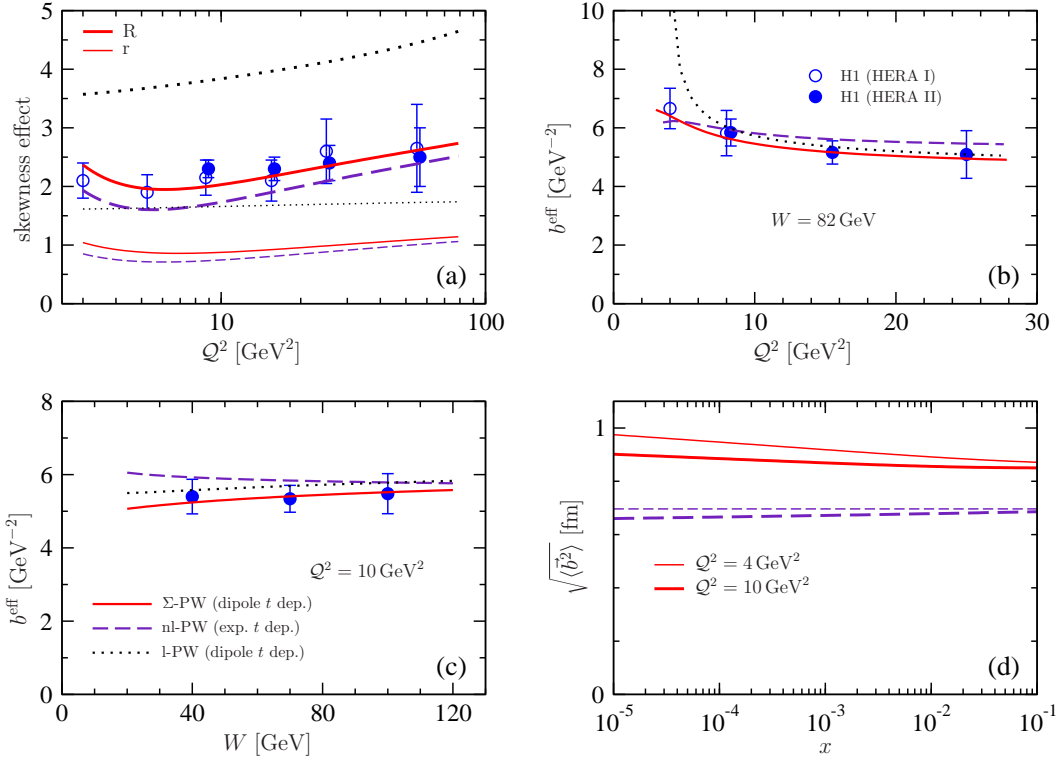}
} \caption{\label{fig:skewslope} (a) the skewness ratio $r$,
cf.~Eq.~(\ref{eq:skewr}), for fixed $x=10^{-3}$ (thin) and
alternative ratio $R$, cf.~Eq.~(\ref{eq:skewR}), for fixed $W=82\,
{\rm GeV}$ (thick), (b) and (c) effective exponential $t$-slope
(\ref{Def-beff}) versus ${\cal Q}$ and $W$, respectively, together
with experimental data from Ref.~\cite{:2007cz}; (d) transverse
width (\ref{Def-TraWid}) of sea quarks at the input
scale ${\cal Q}^2=4\,\mbox{GeV}^2$ (thin) and evolved at ${\cal
Q}^2=10\,\mbox{GeV}^2$ (thick). The model parameter for l (dotted), nl (dashed),
and $\Sigma$ (solid) SO(3)-PW models, see text, are
obtained from LO fits such as in Fig.~\ref{fig:fit}. }
\end{figure}
We define a \emph{skewness} ratio of singlet quark GPD on the
$\eta=x$ and $\eta=0$ trajectories:
\begin{equation}
r(x,{\cal Q}^2) \equiv \frac{{}^{\Sigma}\!H(x,\eta=x,t=0,{\cal Q}^2)}
{{}^{\Sigma}\!H(x,\eta=0,t=0,{\cal Q}^2)} \;.
\label{eq:skewr}
\end{equation}
Resulting from our LO fit in Fig.~\ref{fig:fit}, $r$ versus
$\mathcal{Q}^2$ is plotted in Fig.~\ref{fig:skewslope}(a) for the
nl- and $\Sigma$-PW model as thin dashed and solid curves,
respectively. One notes that for both models $r\sim 1$ over the wide
${\cal Q}^2$ lever arm, i.e., a skewness effect is almost absent
at LO and the models are not distinguishable. In these models the
gluonic $r$ ratio is considerably smaller than one and can even be
negative. The ratio $R$ (thick curves) of imaginary parts of DVCS
amplitude and DIS structure function might be considered as an
observable \cite{:2007cz}. This alternative measure of the
skewness effect in the LO approximation reads
\begin{equation}
R\approx \frac{H(x,x,t=0,{\cal Q}^2)}{H(2x,0,t=0,{\cal Q}^2)}.
\label{eq:skewR}
\end{equation}
Note that $R \approx 2^{\alpha(0,{\cal Q}^2)}\, r$ with an
intercept $\alpha(0,{\cal Q}^2\sim 4\,{\rm GeV}^2) \sim 1.2$.
Numerous GPD models and the claim $r\sim 1.5$
\cite{ShuBieMarRys99}, which is implemented in the l-PW model
(dotted curves) and arises from the Clebsch-Gordan coefficient
(\ref{CG-coe}), are disfavored by our LO fits. As mentioned above,
we are even concerned about their uses in fixed target kinematics,
if described at LO \cite{GuzTec08}. We emphasize that l-PW models
with $r\sim 1.5$ can describe the DVCS data beyond LO
\cite{Kumericki:2007sa}, where the corresponding gluonic ratio is
about one.

The $t$-dependence of the DVCS data can be directly fitted within
an exponential ansatz, where a ${\cal Q}$ dependent $t$-slope and
a vanishing $\alpha^\prime$  has been found \cite{:2007cz}. We
display in Fig.~\ref{fig:skewslope} (b) and (c) the H1
measurements \cite{:2007cz} and the effective exponential
$t$-slope
\begin{eqnarray}
\label{Def-beff}
b^{\rm eff} = \frac{1}{-0.7\, {\rm GeV}^2}
\ln\frac{
\frac{d\sigma_{\rm DVCS}}{dt}(W,t=-0.8 \,{\rm GeV}^2,{\cal Q}^2)
}{
\frac{d\sigma_{\rm DVCS}}{dt}(W,t=-0.1 \,{\rm GeV}^2,{\cal Q}^2)}\,,
\end{eqnarray}
evaluated from  a $\Sigma$-PW model (solid) with dipole ansatz and
$\alpha^\prime=0.15/{\rm GeV}^2$ as well as a nl-PW model (dashed)
with exponential ansatz and $\alpha^\prime=0$. Models describe the
present DVCS data set with $\chi^2/{\rm d.o.f.}\approx 101./98$
and $\chi^2/{\rm d.o.f.}\approx 98./98$, respectively. We
emphasize that the differences between the nl- and $\Sigma$-PW
model, visible in Fig.~\ref{fig:skewslope}, mainly originate from
the implementation of $t$-dependence. The decrease of the
$t$-slope with growing photon virtuality ${\cal Q}$, shown in
panel (b), {\it entirely} arises from evolution. The flatness of
the $t$-slope with respect to the $W$-dependence, see panel (c),
does not exclude a small $\alpha^\prime$ value. Note that due to
the evolution the $\alpha^\prime$ value at the scale ${\cal
Q}^2=10\, {\rm GeV}^2$ is smaller than the input value
$\alpha^\prime=0.15/\mbox{GeV}^2$. We add that from our fits a
large value of $\alpha^\prime \sim 0.8/\mbox{GeV}^2$ for the
dominant sea quark contribution is disfavored, cf.~Ref.~\cite{GuzTec08}.

Present small-$x_{\rm Bj}$ DVCS data are well described by different
GPD models and thus a partonic interpretation w.r.t.~the skewness
dependence, i.e., the distribution of SO(3) partial waves, and the
$t$-dependence can not be unique. For the transverse width
\cite{Bur02}
\begin{eqnarray}
\label{Def-TraWid}
\langle \mbox{\boldmath  $b$}^2 \rangle(x,{\cal Q}^2) =
4 \frac{d}{d t} \ln H(x,\eta=0,t,{\cal Q}^2)
\end{eqnarray}
of sea quarks this is illustrated in Fig.~\ref{fig:skewslope} (d).
Here the differences for the transverse width of the dipole
(solid) and exponential (dashed) $t$-dependent model are caused by
their extrapolation descriptions of experimental data to $t=0$. We
also show the evolution of the transverse width from ${\cal
Q}^2=4\,\mbox{GeV}^2$ (thin) to ${\cal Q}^2=10\,\mbox{GeV}^2$
(thick). Thereby, the slope $\alpha^\prime$ decreases, compare
thick and thin solid curves.

We remark that our models also reproduce the preliminary H1 result
on the dominant twist-two $\cos\phi$ harmonic for the beam charge
asymmetry \cite{Sch07}. Since this asymmetry is proportional to
the real part of a CFF combination,  its description  gives us
confidence in the Regge behavior of the DVCS amplitude.
Unfortunately, we could not obtain a useful
bound for the  CFF $\cal E$ or the related  anomalous gravitomagnetic
moment, which would shed light on the partonic  decomposition of
the proton spin.

{\ }

\noindent
In conclusion, we successfully describe the small $x_{\rm Bj}$
DVCS data within flexible GPD models from the perspective of
$t$-channel physics or Regge phenomenology. Thereby, a LO
description requires the inclusion of non-leading SO(3) partial
waves, while beyond this order they are not essential. The
interplay of both $t$ and $\xi$ dependence with ${\cal Q}^2$
evolution is in our GPD models compatible with data. In a partonic
LO interpretation the dominant sea quark GPD possesses almost no
skewness effect over a wide ${\cal Q}^2$ lever arm. This feature
can be realized within various models. We also found that gluons,
driving the evolution at small $x$, are relatively suppressed
(even a negative gluonic GPD on the trajectory $\eta=x$). The
functional form of the $t$-dependence in the GPD models can not be
fully pinned down from present DVCS data.  In particular, a small,
however, non zero, $\alpha^\prime$ can not be excluded and the
residual $t$-dependence might be given by a dipole ansatz. For a
global fit of DVCS data, a good LO description of small-$x_{\rm
Bj}$ DVCS data is valuable at present. Whether one prefers a quark
interpretation of DVCS data in a specific factorization scheme, in
which absence of skewness effect holds at any order, or likes to
resolve the gluonic content within radiative corrections in the
standard scheme, is primarily a convention, yielding a GPD
reparameterization.

\section*{\large Acknowledgments}
This work was supported in part by the BMBF (Federal Ministry for
Education and Research), contract FKZ 06 B0 103 and by the
Croatian Ministry of Science, Education and Sport under the
contracts no. 119-0982930-1016 and 098-0982930-2864.


\begin{footnotesize}

\end{footnotesize}


\end{document}